\documentclass[12pt]{iopart}
\usepackage{graphicx}
\begin{document}
\paper[]{Immersed nano-sized Al dispersoids in an Al matrix; 
effects on the structural and mechanical properties by 
Molecular Dynamics simulations}
\author{H Chamati\footnote{Permanent address: 
Institute of Solid State Physics, 
72 Tzarigradsko Chauss\'ee, 1784 Sofia, Bulgaria},
M S Stoycheva\footnote{Permanent address: 
Geophysical Institute, Acad. G. Bonchev str, 
bl.3, 1113 Sofia, Bulgaria}
and G A Evangelakis}
\address{Department of Physics, Solid State Division, 
University of Ioannina, P.O. Box 1186, GR-45110 Ioannina, Greece}
\eads{\mailto{chamati@issp.bas.bg}; \mailto{gevagel@cc.uoi.gr}}

\begin{abstract}
We used molecular dynamics simulations based on a potential model in
analogy to the Tight Binding scheme in the Second Moment Approximation
to simulate the effects of aluminum icosahedral grains (dispersoids) on
the structure and the mechanical properties of an aluminum matrix. First
we validated our model by calculating several thermodynamic properties
referring to the bulk Al case and we found good agreement with available
experimental and theoretical data. Afterwards, we simulated Al systems
containing Al clusters of various sizes. We found that the structure of
the Al matrix is affected by the presence of the dispersoids resulting
in well ordered domains of different symmetries that were identified
using suitable Voronoi analysis. In addition, we found that the increase
of the grain size has negative effect on the mechanical properties of
the nanocomposite as manifested by the lowering of the calculated bulk
moduli. The obtained results are in line with available experimental
data.
\end{abstract}
%\begin{keyword}
%interatomic potentials; elastic constants;
%molecular-dynamics simulations; nanostructures; nanocomposites;
%mechanical properties.
%\end{keyword}

\submitto{\JPCM}

\maketitle

%\newpage

\section{Introduction}
Nanostructuring is usually used to improve the mechanical properties of
bulk (coarse-grained) metals and alloys. In this context structures in
the nanometer range as precipitates, different phases or dislocations
arrays are introduced into the materials. The reduction of grain size
down to the nanometer regime has opened new and fascinating avenues for
research in several aspects of materials science. Nanocrystalline
materials are believed to exhibit quite different responses under
irradiation environments compared to coarser-grained materials. From
theoretical point of view it is interesting to model nanocrystalline
metals since the small grain size results in a cutoff of the typical
length scale of the phenomena and structures that may appear during the
deformation process. For a further discussion of mechanical properties
of nanocrystalline metals see References
\cite{morris1997,hauert2000,cheng2003} and references therein.

The investigation of the physical as well as chemical properties of 
metallic clusters received considerable attention in the last few 
decades (for a review on the experimental studies see Reference 
\cite{marks1994}). The size of these clusters is typically ranging
from 1-100 nanometers, whence the name `nanoclusters' is used for these
particles. Because of their properties that are very different
from their bulk or nanocrystalline counterparts they are very promising
in many technological applications. Usually they adopt structures that
are completely different from their bulk parents. It
has been found both experimentally and analytically that they may exist in
many different structures: icosahedral, decahedral, Marks polyhedra, etc.
The change in the behaviour of these nanosized particles is originating
mainly from their large surface to volume ratio, while the major
contributions to the physical and chemical properties originate from their
surfaces.

The interest in materials that are an assemblage of nanometer-sized
particles, the so called nanocomposite, arises from the realization that by
controlling their size we can alter a variety of
physical, mechanical and chemical properties of the bulk materials (see
Reference \cite{murthy2003} and references therin).

According to the physical theory of strength the dispersion of
crystalline structures inside a polycrystalline matrix promotes
essential rise of strengthening and improves low temperature plasticity
of the materials. Moreover, microcrystalline materials can exhibit
superplasticity at high temperatures.  High mechanical properties could
be expected when extrapolating this tendency to nanocrystalline
structures; however, as the amount of experimental data increases, it
becomes obvious that only a few studies confirm such an assumption.

Small nano-sized particles dispersed homogeneously throughout the bulk
of an alloy enhance its strength by impeding dislocation motion
\cite{hauert2000}. The dispersion of such grains is achieved either by
mechanical mixing with the matrix material or by precipitation from a
supersaturated solution. The former provides independent control of
the size, type, morphology and volume fraction of the grains, called
dispesoids. The effectiveness of `dispersoids' depends on their size,
spacing and distribution. The characteristics of the interface
separating the dispersoid and the matrix play a crucial role on the
structural and mechanical properties of the nanocomposites.

In the present work we investigate the effects of Al dispersoids on an
Al matrix. A nano-sized grain was placed at the center of a liquid Al
matrix. Using Molecular dynamics we investigated the structural as well
as the mechanical alterations of the Al matrix compared to Al bulk
properties.

The paper is organized as follows. In Section \ref{intro} we present
some thermodynamic quantities of the bulk aluminum along with
the corresponding experimental data. In Section \ref{effects} we
describe the dispersoids and we investigate their effects in a bulk
aluminum matrix, to end up with a discussion that is given in Section
\ref{discussion}.

\section{Physical properties of bulk aluminum}\label{intro}
It is known that interatomic potentials play a key role in any
computer simulation in materials science. Much progress has been done
on the development of semi-empirical interatomic potentials including
the so-called `many-body interactions'. These terms are meant to take
into account the local electronic density. Among several many-body
potentials, the most popular are the embedded atom model (EAM), the
Finnis-Sinclair potentials (FS) and the tight-binding potential in the
second moment approximation. Though based on different approaches these
models are shown to be equivalent \cite{voter1994} at least from the
mathematical point of view.

In the present work, we choose a tight-binding potential in the second
moment approximation, according to which the total energy of the system
can be expressed as:
\begin{equation}\label{energy}
E=\sum_i\left[\sum_{j\neq i}Ae^{-p(r_{ij}/r_0-1)}
-\xi\sqrt{\sum_{j\neq i}^{}e^{-2q(r_{ij}/r_0-1)}}\right],
\end{equation}
where the first sum over $j$ is a Born-Mayer type pairwise potential
adapted for the description of the repulsive contributions, while the
second one represents the band-structure term. In the above expression
(\ref{energy}), $r_{ij}$ is the distance between atoms $i$ and $j$,
while the interactions up to fifth neighbors were taken into account.
The adjustable parameters $A$, $\xi$, $p$, $q$ and $r_0$ in this
approach have been determined through fitting to the total energy of the
system as computed by first-principles APW calculations and taking into
account the experimental values of the cohesive energy, the lattice
constant and the structural stability of the ground state (see details
in Reference~\cite{papanicolaou2003}). The values of the parameters are
given in Reference~\cite{papanicolaou2003}: $A=0.05504$ eV, $\xi=0.9564$
eV, $p=10.9011$, $q=1.5126$ and $r_0=2.8310$ \AA. We note that this
model is composed of exponential functions resulting in high numerical
reproducibility.

As a test of the quality of our parameters, in a previous paper
\cite{papanicolaou2003}, we calculated some bulk properties: bulk
moduli, elastic constants, phonon density of states at room temperature,
temperature dependence of the lattice constant as well as the mean
square-displacement. The obtained results were found to be in good 
agreement with the corresponding experimental values.

Using the above interatomic potential, we performed additional molecular
dynamics simulations in the isothermal canonical ensemble (NVT) using
the NOSE deamon to control the temperature and the Verlet algorithm with
a time step of 5 fs for the integration of the equations of motion. The
system was made up of 4000 particles arranged in the appropriate lattice
structure.

\begin{table}
\caption{ Computed properties of bulk Al along with experimental, 
{\it ab initio} data, and EAM potential results 
~\cite{mishin1999}}\label{table1}
\begin{center}
\begin{indented}
\begin{tabular}{lccc}
\hline
\hline 
               & Experiment         & Present        & EAM             \\
               & or {\it ab initio} & work           & 
	                                       (Reference \cite{mishin1999})\\
\hline
\underline{Lattice properties:} \\
$a_0$(\AA)               &   4.05$^a$  &    3.995      &    4.05       \\
$E_0$(eV/atom)           &  -3.36$^b$  &    -3.34      &   -3.36       \\
$B(10^{12}Pa)$           &   0.79$^c$  &    0.73       &    0.79       \\
$c_{11}(10^{12}$Pa)      &   1.14$^c$  &    0.92       &    1.14       \\
$c_{12}(10^{12}$Pa)      &   0.619$^c$ &    0.63       &    0.616      \\
$c_{44}(10^{12}$Pa)      &   0.316$^c$ &    0.48       &    0.316      \\
\underline{Phonon frequencies:} \\
$\nu_L(X)$(THz)            &    9.69$^d$ &    9.70       &    9.31       \\
$\nu_T(X)$(THz)            &    5.80$^d$ &    6.25       &    5.98       \\
$\nu_L(L)$(THz)            &    9.55$^d$ &    9.52       &    9.64       \\
$\nu_T(L)$(THz)            &    4.19$^d$ &    4.10       &    4.30       \\
$\nu_L(K)$(THz)            &    7.59$^d$ &    7.45       &    7.30       \\
$\nu_{T_1}(K)$(THz)        &    5.64$^d$ &    5.45       &    5.42       \\
$\nu_{T_2}(K)$(THz)        &    8.65$^d$ &    8.40       &    8.28       \\
\underline{Vacancy:} \\
$E_v^f$(eV)              &    0.65$^e$ &    1.28       &    0.68       \\
\underline{Surfaces:} \\
$\gamma_s$(110)(mJ/m$^2$)&    1200$^f$ &    1410       &    1006       \\
$\gamma_s$(100)(mJ/m$^2$)&    1200$^f$ &    1320       &    943        \\
$\gamma_s$(111)(mJ/m$^2$)&    1200$^f$ &    1290       &    870        \\
\hline
\hline \\
\end{tabular}
\end{indented}
\end{center}

$^a$ Reference \cite{kittel1986}

$^b$ Reference \cite{weast1984}

$^c$ Reference \cite{smith1976}

$^d$ Reference \cite{simons1977}

$^e$ Reference \cite{tyson1977}

$^f$ Reference \cite{schaefer1987}
\end{table}

Here we present further results (see Table~\ref{table1}) computed
within the framework of this model. A comparison of our results with
experimentally available values shows that our model reproduces quite
well the equilibrium properties, elastic constant (except C$_{44}$) as
well as the surface energies. The elastic constants are given here for
sake of completeness.

\begin{figure}
\centerline{
\resizebox{5.0in}{!}{\includegraphics{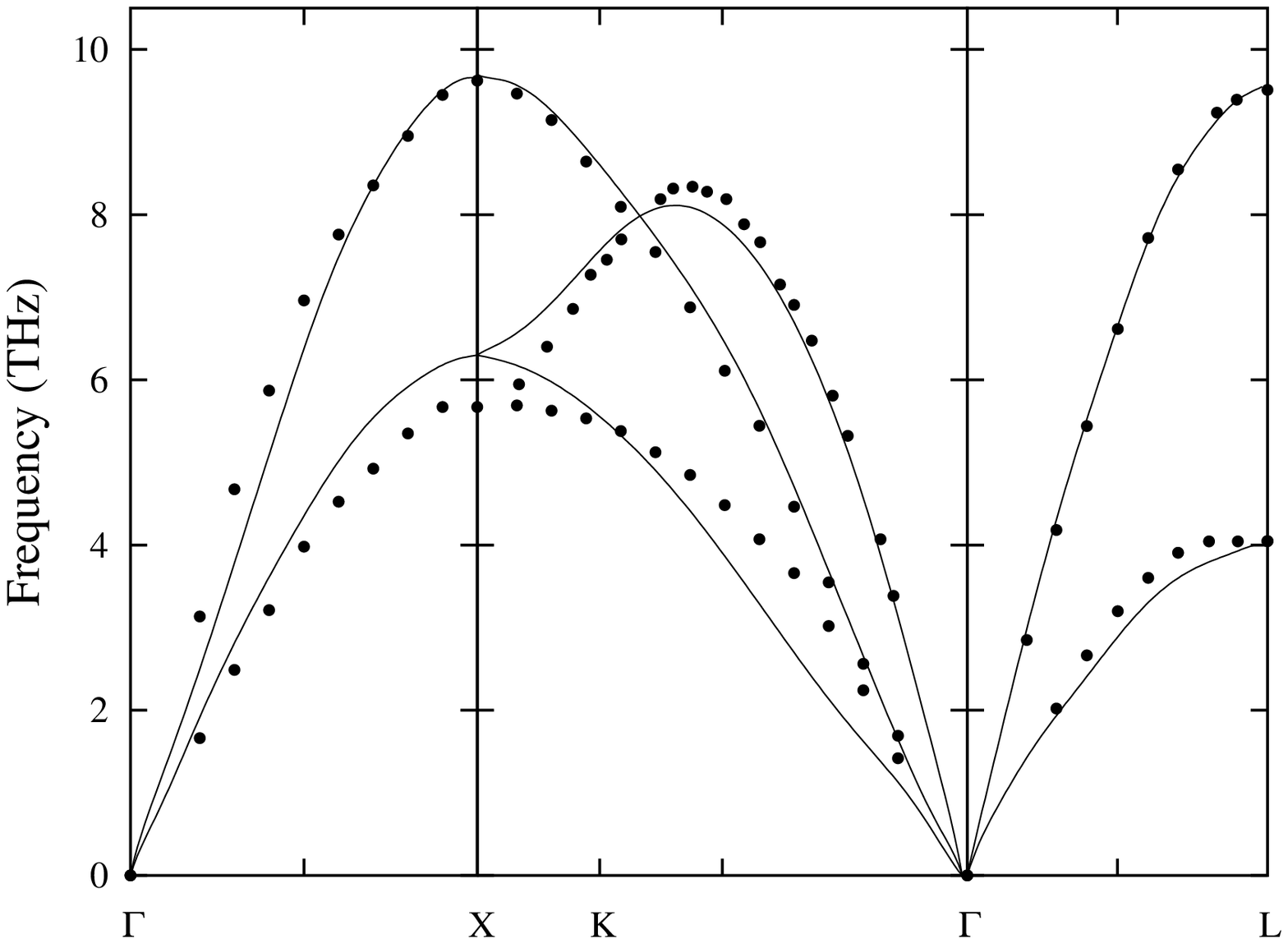}}}
\vspace{.3cm}
\caption{Phonon dispersion curves of Al at 300K. Solid lines correspond 
to molecular dynamics simulations, filled circles refer to 
experimental data from Reference~\protect\cite{stedman1966}.}
\label{disp}
\end{figure}

In Figure \ref{disp}, we present the phonon-dispersion curves for aluminum
at room temperature, along with the corresponding neutron scattering
data \cite{stedman1966}. We observe a remarkable accuracy of our model
for the high-symmetry branches and the cutoff frequency except for a
slight overestimation of about 1 THz for the transverse mode along [100]
and T$_1$, and L along [110] direction, in the vicinity of the point X.
This is compatible with the inaccuracy found in our calculated value of
C$_{44}$.

In addition, we calculated the surface energies for the three low
indexed surfaces. The values we obtained are slightly higher than the
experimental data \cite{tyson1977}, a result that is opposite from what
it is usually found using EAM type potentials (see
Reference~\cite{mishin1999} for more details), but in reasonable
agreement with the experimental and theoretical data. On the contrary,
the evaluation of the vacancy formation energy yielded a value that is
not close to the experimental data. This result has to be attributed to
the charge transfer between neighbors around the vacancy, an effect that
requires different approach in order to be treated properly; in our
model the electronic contributions are incorporated into the second term
of the potential energy functional and therefore cannot be used to take
into account eventual electronic modifications.

Moreover we estimated the melting temperature of our potential model for
aluminum using the Lindemann's criterion \cite{lindemann1910}. According
to this criterion, at the melting temperature the average amplitude of
vibration is about 8\% of the nearest-neighbor distance. A linear
extrapolation of our atomic mean-square displacement results
\cite{papanicolaou2003} gives a melting temperature of 960 K, consistent
with the experimental value of 933 K of high-symmetry crystal
structures.

Turning our attention to the mechanical properties, we calculated the
elastic constants since they yield valuable information concerning the
strength and the stability of materials. A knowledge of their
behaviour over a wide range of temperature is of a fundamental importance
in characterising a large number of mechanical, electric, optical and
thermodynamic properties. The molecular dynamics computation of
isothermal elastic constants, $c^T_{ijkl}$, is given by the following
expressions \cite{squire1969,zhou2002}
\begin{eqnarray*}
c_{11}^T&=&
\frac1{Vk_BT}\left(\langle\varphi_{ii}\rangle^2
-\langle\varphi_{ii}^2\rangle\right)
+\frac1V\langle\varphi_{}\rangle+\frac{2Nk_BT}V\\
c_{12}^T&=&
\frac1{Vk_BT}\left(\langle\varphi_{ii}\varphi_{jj}\rangle
-\langle\varphi_{ii}\varphi_{jj}\rangle\right)
+\frac1V\langle\varphi_{iijj}\rangle\\
c_{44}^T&=&
\frac1{Vk_BT}\left(\langle\varphi_{ij}\rangle^2
-\langle\varphi_{ij}^2\rangle\right)
+\frac1V\langle\varphi_{ijij}\rangle+\frac{Nk_BT}V,
\end{eqnarray*}
where $V$ is the volume of the simulation box and $k_B$ the Boltzmann
constant; $T$ is the temperature; $\varphi_{ij}$ and $\varphi_{ijkl}$
are the first and the second derivatives of the potential energy
$\varphi$ with respect to the strain tensor components $\eta_{ijkl}$
respectively and the brackets denote canonical averages. Thermodynamic
averages were obtained from equilibrium molecular dynamics trajectories
that lasted 50 000 time steps.

\begin{figure}
\centerline{
\resizebox{5.0in}{!}{\includegraphics{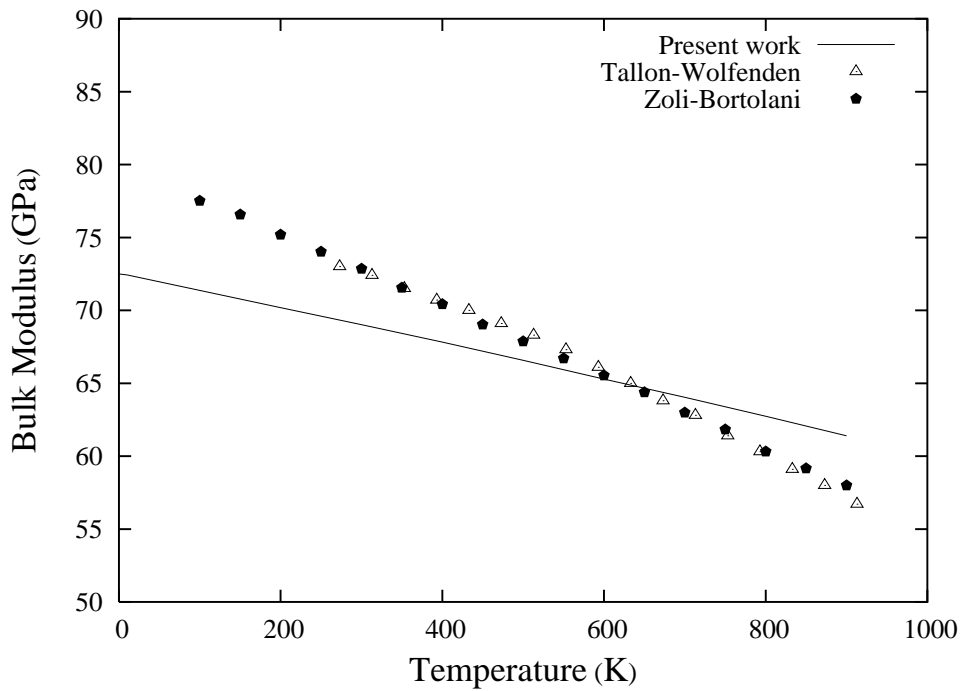}}}
\caption{Temperature dependence of the isothermal bulk modulus of
aluminum. The experimental results are taken from
Reference \cite{tallon1979} and the theoretical 
from Reference \cite{zoli1990}.}
\label{bulk}
\end{figure}

Here we present only results referring to the temperature dependence
of the bulk modulus (Figure \ref{bulk}). Note that the values of this
thermoelastic quantity were obtained using the relationship
$B=(c_{11}+2c_{12})/3$ valid for cubic lattices \cite{mehl1994}. We found a
fairly good agreement with experimental values \cite{tallon1979} and
theoretical calculations \cite{zoli1990}.

It comes out that the present model reproduces quite satisfactorily the
main features of the bulk Al and it can therefore be used for 
simulations of at least the characteristic properties of aluminum
nanostructures.

\section{Effects of nano-Al dispersoids on the bulk system}
\label{effects}
\subsection{Molecular dynamics procedure}
A criterion that is helpful in finding the most stable configuration of
a dispersoid (nanocluster) is the use of the Wulff construction that was
developed for finding the equilibrium shape of a quasi-crystal made of a
given number of atoms by minimizing its surface energies. In the present
study the number of atoms has been chosen to correspond to the magic
numbers identifying icosahedra \cite{martin1996}. These are structured
in shells. An icoshedra with $k$ shells has
\begin{equation}\label{magic}
N(k)=\frac{10}3k^3-5k^2+\frac{11}3k-1
\end{equation}
atoms (so the series of magic numbers is 1, 13, 55, 147, \ldots) 
exhibiting 20 triangular facets of side $k$ (for more details see 
\cite{martin1996}).

Using different approaches, various aluminum cluster sizes (Al$_n$,
where $n$ indicates the number of atoms composing the cluster) are
investigated in the literature. It has been found using an {\it ab
initio} study \cite{yi1991} that the Al$_{55}$ cluster presents
substantial structural distortions in its lowest-energy configuration
with respect to the highly symmetric icosahedral. Several inequivalent
and energetically degenerated structures were found. The origin of this
degeneracy is traced to the short range effective interatomic potential.
Using molecular dynamics investigation combined with a simulated
annealing technique, Reference \cite{sun1997} gave results referring to
the structural and dynamical properties of Al$_n$ clusters
($n=13,55,\textrm{ and } 147$) that were found to agree with {\it
ab-initio} calculations.  Very recently the electronic and thermodynamic
properties of Al$_n$ clusters ($n=55,147,\textrm{ and } 309$) were
determined. It was found that these clusters have melting points and
bulk moduli lower than those corresponding to the  bulk Al, while they
exhibit enhanced low and high energy phonon densities of states
\cite{mitev2003}. The calculated electronic density of states revealed
significant enhancement at energies around the Fermi level, indicating
clear quantum confinement.  This is due to the charge transfer from the
inner atoms towards the surface atoms \cite{mitev2003}.

To obtain the lowest-energy configuration of each cluster we used a
simulated annealing and quenching technique. These techniques are the
best available procedures to find the global and local minima in
complicated situations. They allow us to test the results by varying
cooling rates and starting configurations. We start with fcc clusters
consisting of a number of atoms corresponding to a magic number (given
by Equation (\ref{magic})) of the icosahedral structure. We heat the
initially constructed clusters up to temperatures that are higher than
the melting point. The resulting liquid is subsequently quenched in
order to minimize the energy. The procedure is repeated several
times and the coordinates of the quenched cluster were saved. Finally,
the cluster with the lowest energy is kept for further study. Let us
emphasize that these clusters are found to have an icosahedral-like
structure as expected.

In order to investigate the effects of the nano-al dispersoids in the Al
matrix, we choose the preparation of a liquid phase by starting with an
fcc lattice corresponding to the aluminum matrix. The lattice constant
has been adjusted to have zero pressure for the liquid at 2000K,
temperature at which the system was thermalized for 20 000 time steps.
Subsequently, we placed five different nano-sized icosahedral (Al$_n$)
grains (dispersoids), consisting of $n=$ 55, 147, 309, 561 and 923
atoms, at the center of the resulting liquid, forming five different
initial configurations of nano-composites. Care has been taken when
placing the grains inside the matrix in order to keep the same
orientation for all grains and be able to compare the effects of grain
size on the matrix eliminating the effect that could possibly be induced
by their different orientations. Next, the samples were cooled down to
500K and allowed to equilibrate for 500 000 time steps. During this time
the dispersoids were fixed at their initial positions. For these final
configurations, we investigated the structural and mechanical
properties.

To facilitate the analysis of the local atomic order of our final
configurations we used the algorithm proposed in Reference
\cite{stankovic2002}. According to this algorithm the bonds between an
atom and its `relevant' neighbors are examined to determine the crystal
structure using a Voronoi construction with self-adjusting cutoff to
recognize the nearest neighbors. This method classifies the atoms into four
classes: bcc, fcc, hcp and icosahedral. Additional details and
advantages of the method can be found in Reference \cite{stankovic2002}.

\subsection{Results}
\begin{figure}
\centerline{
\resizebox{5.4in}{!}{\includegraphics{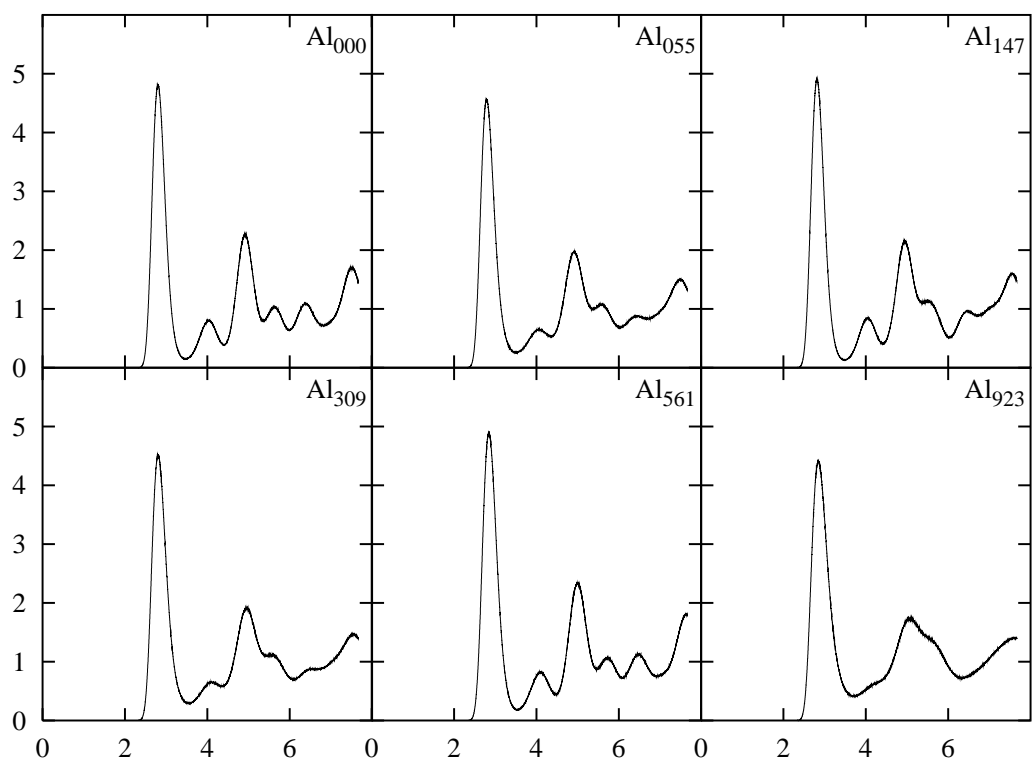}}}
\caption{Radial distribution functions of the equilibred aluminum 
systems containing various dispersoids}
\label{pdf}
\end{figure}

\begin{figure}
\centerline{
\resizebox{5.4in}{!}{\includegraphics{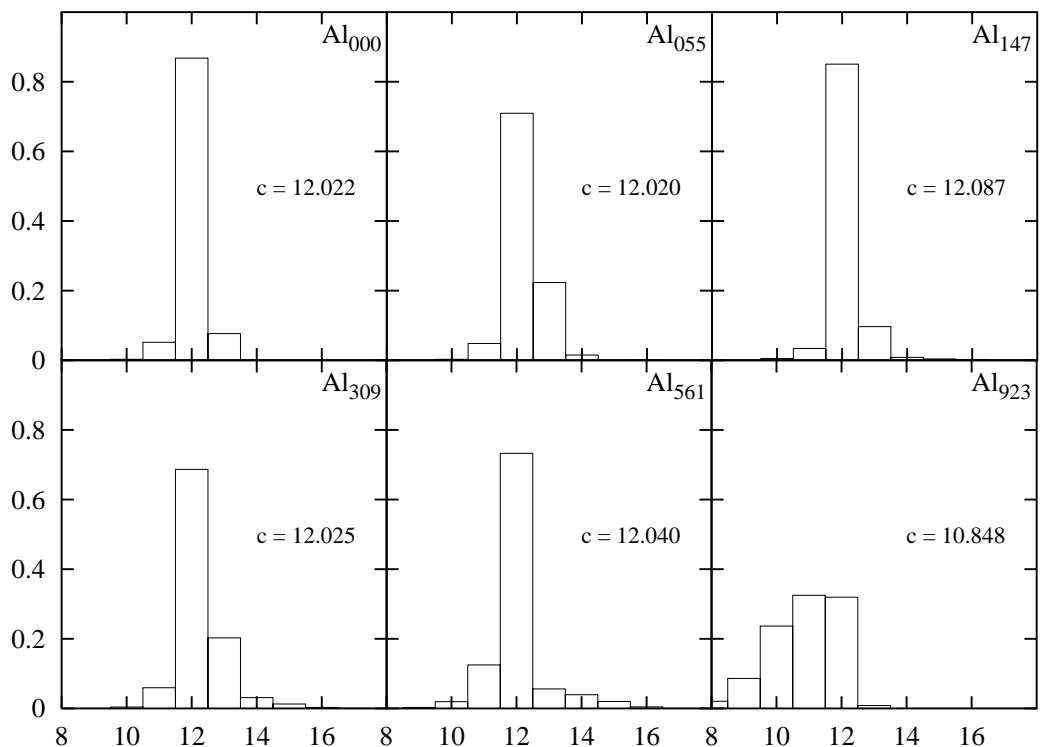}}}
\caption{Coordination histograms obtained from Voronoi structural 
analysis of the final configurations. The
quantity c is the mean coordination number.}
\label{voronoi}
\end{figure}

In Figure \ref{pdf} we present the average pair distribution functions
(obtained from the molecular dynamic runs) of the various Al samples
with grains inside. The index $n$ in Al$_n$ indicates the number of
atoms forming the dispersoids. The configuration labeled as Al$_{000}$
corresponds to the free of grains case and it serves as a reference
configuration. The peaks could be related one by one to the different
coordination shells of the polyclystalline matrix. We observe that the
grain size affects the structure of the equlibrated aluminum matrix at
500K. More specifically, the reference matrix (Al$_{000}$) has more
pronounced peaks than the others and its pair distribution function is
more similar to the fcc structure.  This is confirmed by the Voronoi
analysis shown in Figure \ref{voronoi}.  By increasing the grain size we
diminish the ratio between the grain size and the aluminum matrix. The
size of the grain affects its structural as well as mechanical
properties. It comes out that although the mean coordination number
remains more or less the same, the number of the atoms having more
neighbors changes. As a result different local structures appear: fcc,
bcc, hcp and icosahedral in different proportions. This is due to the
embedded icosahedral grains. The largest grain i.e. the one with 923
atoms affects the system more drastically leading to an (amorphous)
glassy structure, while the coordination number in this case is lower
than in the cases with smaller grains.

\begin{figure}
\centerline{
\resizebox{5.0in}{!}{\includegraphics{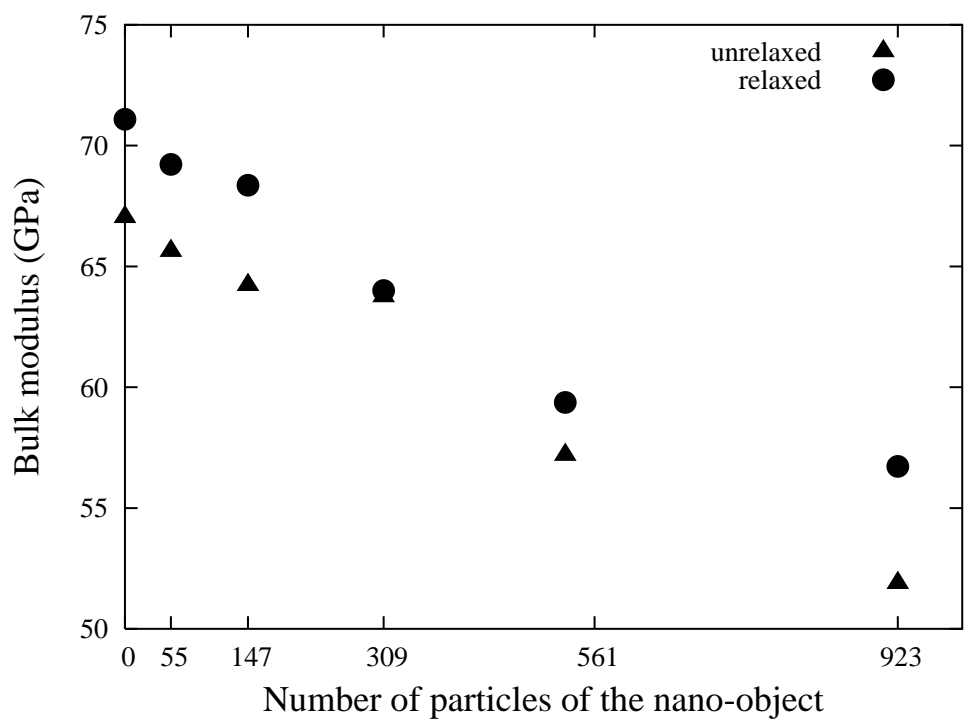}}}
\caption{Static bulk moduli of the various final configurations.}
\label{nanobulk}
\end{figure}

The effects of the grain size on the mechanical properties of the
polycrystalline reference configuration have been verified by
calculating the bulk modulus of the final configurations. We consider
that this quantity can be used to obtain an evidence of the effect of
the grain size on the mechanical properties of the system, at least
qualitatively. The bulk modulus is evaluated using
\begin{displaymath}
B=V\frac{\partial^2 E}{\partial V^2},
\end{displaymath}
expressing the variation of the potential energy $E$ with respect to the
volume $V$ of a given `static' configuration at the minimum.  The
calculations were carried out on the final configurations obtained from
the quenched equilibrated systems at 500K. The results are depicted in
Figure \ref{nanobulk}. We see that the bulk modulus decreases as the
grain size increases, indicating softening of the matrix, in agreement
with experimental observations \cite{bonetti1997}.

\begin{figure}
\begin{center}
\resizebox{5in}{!}{\includegraphics{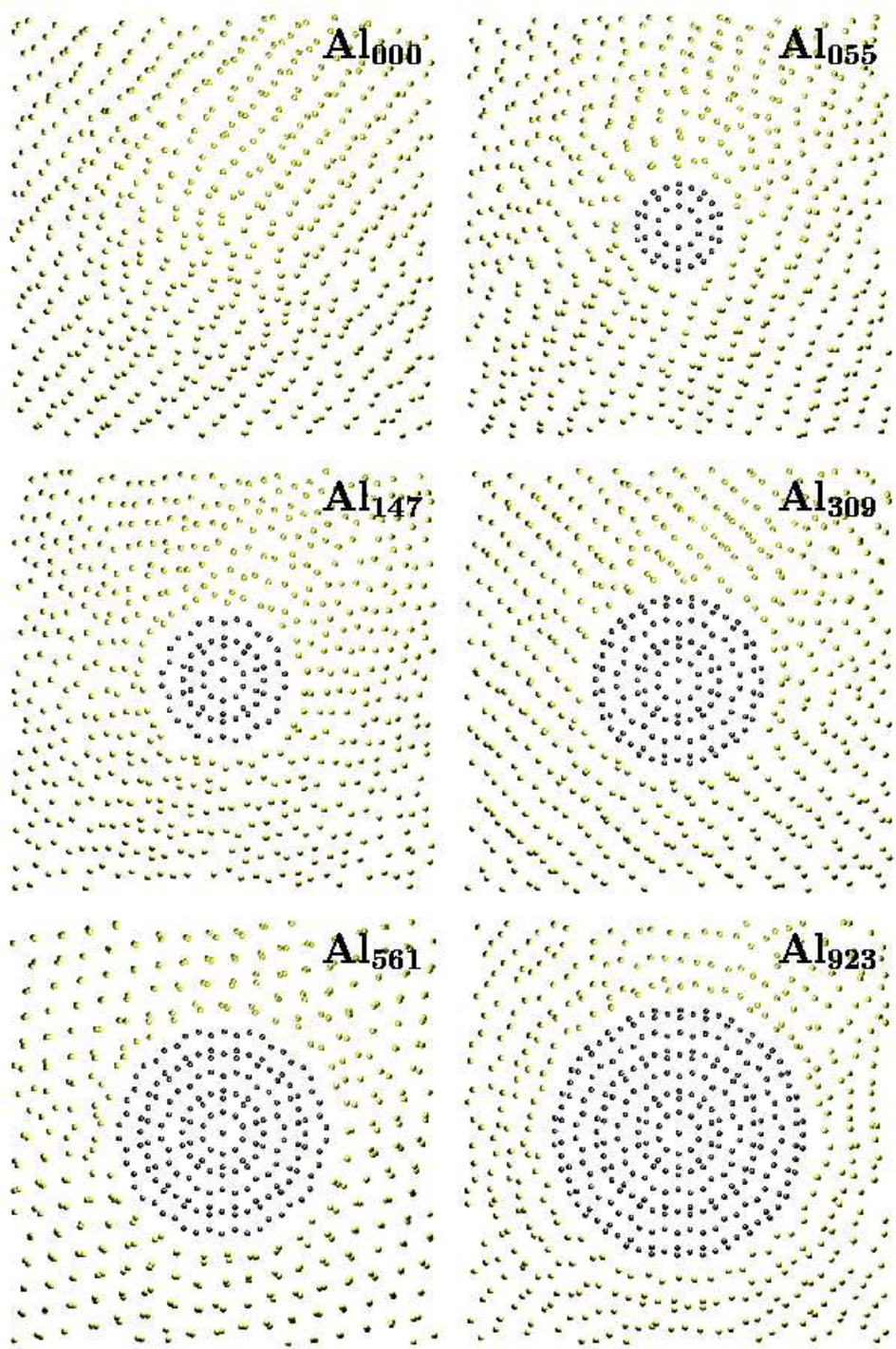}}
\caption{Selected snapshots of the final configurations showing the
dispersoids inside the simulated systems.}
\label{snapshots}
\end{center}
\end{figure}

The final configurations we used are shown in Figure \ref{snapshots}. We
observe that the atoms surrounding the grains order following their
icosahedral structure, while the atoms far from the grains form
different local structures in agreement with the Voronoi analysis
described above. We note also that the atoms around the grains form
icosahedral shells, increasing in this way the original size of the
dispersoid.

\begin{figure}
\centerline{
\resizebox{5in}{!}{\includegraphics{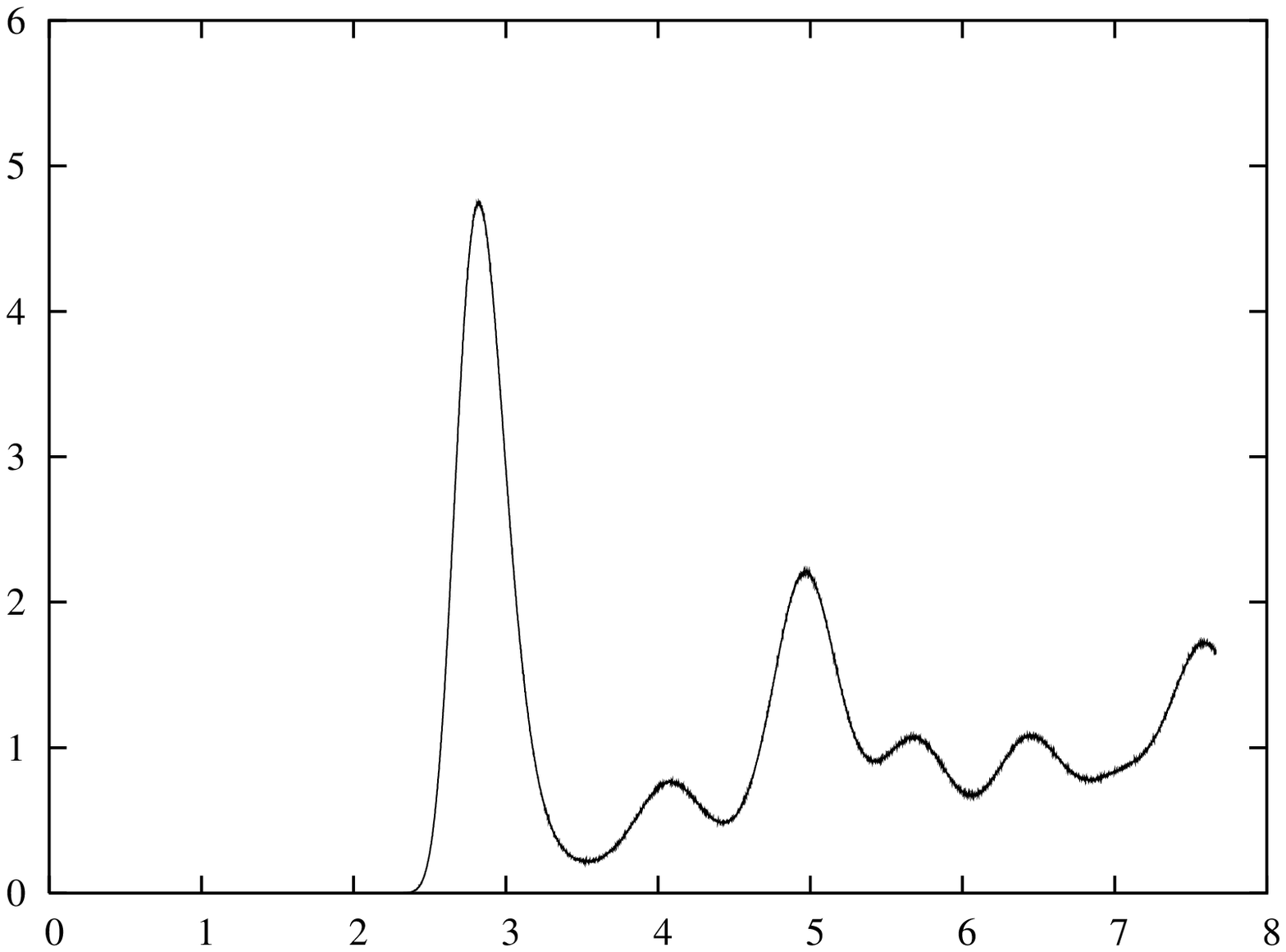}}}
\caption{The same as in Figure~\ref{pdf} for the large system.}
\label{pdflarge}
\end{figure}

\begin{figure}
\centerline{
\resizebox{5in}{!}{\includegraphics{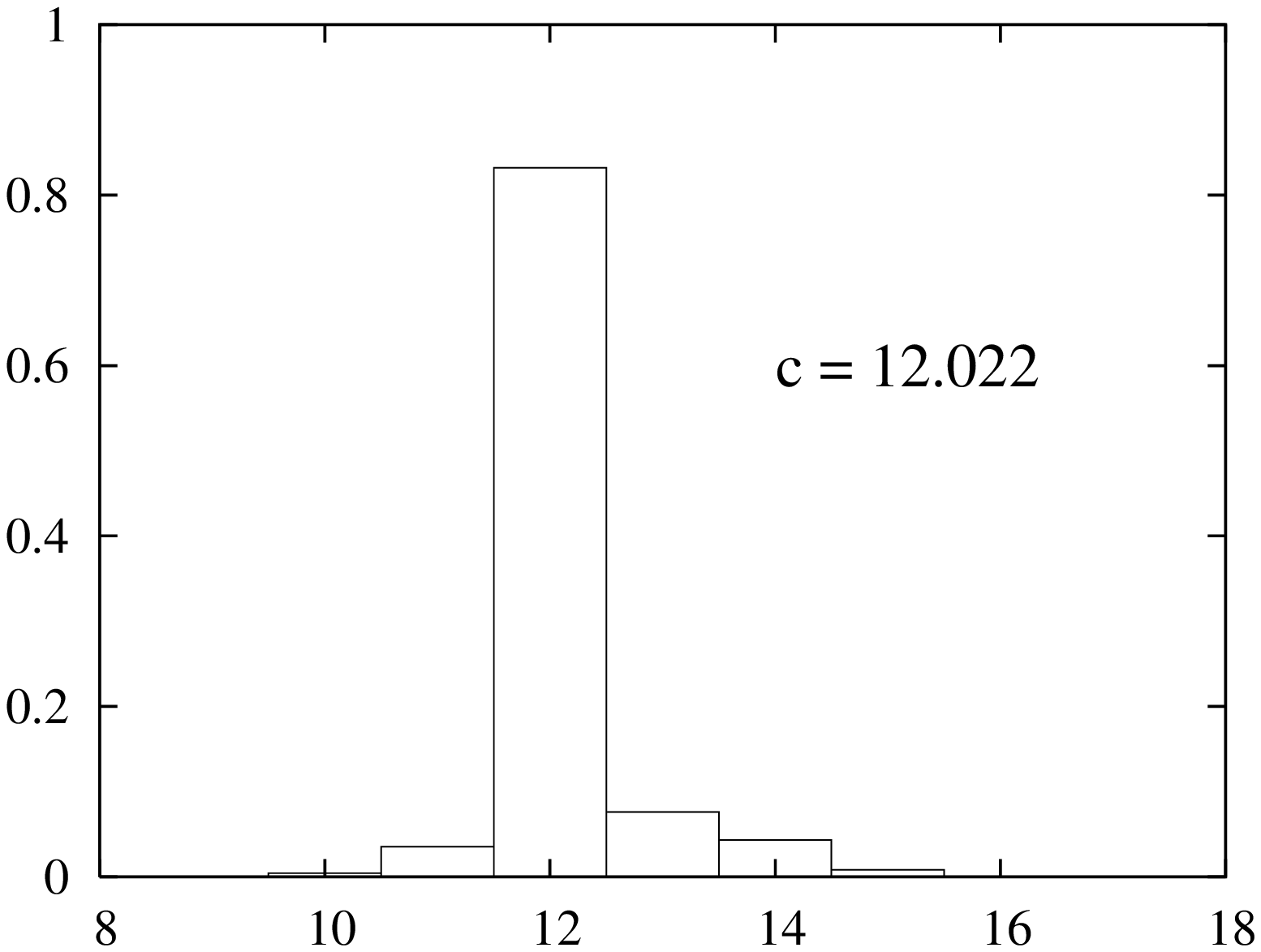}}}
\caption{The same as Figure~\ref{voronoi} for the large system.}
\label{vorlarge}
\end{figure}

In order to test whether our results depend on the size of the cell and
the simulation time we performed molecular dynamics using a large system
of 10976 atoms and an icosahedral nanoparticle of 923 atoms immersed at
its center. The corresponding pair distribution function, Figure
\ref{pdflarge}, exhibits the characteristic peaks of an fcc structure,
while the Voronoi analysis, given in Figure \ref{vorlarge}, reveals that
the system orders in locally different structures. The situation is
similar to the system of 4000 atoms with Al$_{309}$. By computing the
bulk modulus we find $B = 61.58$ GPa for the unrelaxed and $B = 64.61$ 
GPa for the relaxed lattices.

\section{Discussion}\label{discussion}
We presented an extensive Molecular Dynamics study of the physical
properties of bulk aluminium and the effect that is induced by nanozised
Al dispersoids immersed in an Al matrix. The simulations were based on
an interatomic potential in analogy to Tight-Binding Scheme in the
second moment approximation. We computed several properties, like bulk
modulus and its temperature dependence, elastic constants, phonon
frequencies, vacancy formation energy and surface energies. The obtained
results show good agreement between simulation and available
experimental data for the bulk case.

Considering that our potential is able to describe well not only the
bulk properties of the fcc aluminum but also other structures, at least
qualitatively, we simulated aluminum matrices containing Al
icosahedral grains of different sizes. From the computed pair
distribution functions we found several peaks that are not
characteristic of any simple structure yielding a system that exhibits a
certain order that is not trivial to be identified. The application of a
Voronoi analysis yielded a clearer picture of the structure of
the aluminum nanocomposites. We found that the system has a polycrystalline
structure with different regions of local order that were
identified to be: fcc, bcc, hcp and icosahedral. The atoms in the
vicinity of the icosahedral grains organize themselves in shells
following the icosahedral structure. This result was confirmed by
visualising the final equilibrium configurations of the molecular dynamics
simulations at finite temperature. We note here that upon releasing the
nanoparticles the system reorders quickly in an fcc like structure.

In addition, we found that the increase of the grain size has negative
effect on the mechanical properties of the nanocomposites as manifested
by the lowering of the calculated corresponding bulk moduli.

Finally we would like to mention that it would be interesting to extend the
present study to other materials, in particular magnetic systems
and to investigate their effects on the aluminum matrix, {\it e.g} 
Nickel dispersoids.

%\section*{Acknowledgments} 
\ack
This work was supported by the European grant HPRN-CT-2000-00038.  

%\newpage

\newpage

\clearpage

\end{document}